\documentclass[12pt]{article}
\usepackage{html}
\usepackage{epsf}
\usepackage{graphicx}
\usepackage{amssymb}
\begin{document}
\title{MATTERS OF GRAVITY, The newsletter of the APS Topical Group on 
Gravitation}
\begin{center}
{ \Large {\bf MATTERS OF GRAVITY}}\\ 
\bigskip
\hrule
\medskip
{The newsletter of the Topical Group on Gravitation of the American Physical 
Society}\\
\medskip
{\bf Number 43 \hfill Winter 2014}
\end{center}
\begin{flushleft}
\tableofcontents
\vfill\eject
\section*{\noindent  Editor\hfill}
David Garfinkle\\
\smallskip
Department of Physics
Oakland University
Rochester, MI 48309\\
Phone: (248) 370-3411\\
Internet: 
\htmladdnormallink{\protect {\tt{garfinkl-at-oakland.edu}}}
{mailto:garfinkl@oakland.edu}\\
WWW: \htmladdnormallink
{\protect {\tt{http://www.oakland.edu/?id=10223\&sid=249\#garfinkle}}}
{http://www.oakland.edu/?id=10223&sid=249\#garfinkle}\\

\section*{\noindent  Associate Editor\hfill}
Greg Comer\\
\smallskip
Department of Physics and Center for Fluids at All Scales,\\
St. Louis University,
St. Louis, MO 63103\\
Phone: (314) 977-8432\\
Internet:
\htmladdnormallink{\protect {\tt{comergl-at-slu.edu}}}
{mailto:comergl@slu.edu}\\
WWW: \htmladdnormallink{\protect {\tt{http://www.slu.edu/colleges/AS/physics/profs/comer.html}}}
{http://www.slu.edu//colleges/AS/physics/profs/comer.html}\\
\bigskip
\hfill ISSN: 1527-3431

\bigskip

DISCLAIMER: The opinions expressed in the articles of this newsletter represent
the views of the authors and are not necessarily the views of APS.
The articles in this newsletter are not peer reviewed.

\begin{rawhtml}
<P>
<BR><HR><P>
\end{rawhtml}
\end{flushleft}
\pagebreak
\section*{Editorial}

The next newsletter is due September 1st.  This and all subsequent
issues will be available on the web at
\htmladdnormallink 
{\protect {\tt {https://files.oakland.edu/users/garfinkl/web/mog/}}}
{https://files.oakland.edu/users/garfinkl/web/mog/} 
All issues before number {\bf 28} are available at
\htmladdnormallink {\protect {\tt {http://www.phys.lsu.edu/mog}}}
{http://www.phys.lsu.edu/mog}

Any ideas for topics
that should be covered by the newsletter, should be emailed to me, or 
Greg Comer, or
the relevant correspondent.  Any comments/questions/complaints
about the newsletter should be emailed to me.

A hardcopy of the newsletter is distributed free of charge to the
members of the APS Topical Group on Gravitation upon request (the
default distribution form is via the web) to the secretary of the
Topical Group.  It is considered a lack of etiquette to ask me to mail
you hard copies of the newsletter unless you have exhausted all your
resources to get your copy otherwise.

\hfill David Garfinkle 

\bigbreak

\vspace{-0.8cm}
\parskip=0pt
\section*{Correspondents of Matters of Gravity}
\begin{itemize}
\setlength{\itemsep}{-5pt}
\setlength{\parsep}{0pt}
\item Daniel Holz: Relativistic Astrophysics,
\item Bei-Lok Hu: Quantum Cosmology and Related Topics
\item Veronika Hubeny: String Theory
\item Pedro Marronetti: News from NSF
\item Luis Lehner: Numerical Relativity
\item Jim Isenberg: Mathematical Relativity
\item Katherine Freese: Cosmology
\item Lee Smolin: Quantum Gravity
\item Cliff Will: Confrontation of Theory with Experiment
\item Peter Bender: Space Experiments
\item Jens Gundlach: Laboratory Experiments
\item Warren Johnson: Resonant Mass Gravitational Wave Detectors
\item David Shoemaker: LIGO Project
\item Stan Whitcomb: Gravitational Wave detection
\item Peter Saulson and Jorge Pullin: former editors, correspondents at large.
\end{itemize}
\section*{Topical Group in Gravitation (GGR) Authorities}
Chair: Daniel Holz; Chair-Elect: 
Beverly Berger; Vice-Chair: Deirdre Shoemaker. 
Secretary-Treasurer: James Isenberg; Past Chair:  Manuela Campanelli;
Members-at-large:
Michael Landry, Nicolas Yunes,
Curt Cutler, Christian Ott,
Kimberly Boddy, Benjamin Farr,
Andrea Lommen, Jocelyn Read.
\parskip=10pt

\vfill
\eject

\vfill\eject

\section*{\centerline
{we hear that \dots}}
\addtocontents{toc}{\protect\medskip}
\addtocontents{toc}{\bf GGR News:}
\addcontentsline{toc}{subsubsection}{
\it we hear that \dots , by David Garfinkle}
\parskip=3pt
\begin{center}
David Garfinkle, Oakland University
\htmladdnormallink{garfinkl-at-oakland.edu}
{mailto:garfinkl@oakland.edu}
\end{center}

David Blair, Neil Cornish, Steven Detweiler, David Kastor, Steven Liebling, and Benjamin Owen have been elected APS Fellows.

Hearty Congratulations!

\section*{\centerline
{100 years ago}}
\addtocontents{toc}{\protect\medskip}
\addcontentsline{toc}{subsubsection}{
\it 100 years ago, by David Garfinkle}
\parskip=3pt
\begin{center}
David Garfinkle, Oakland University
\htmladdnormallink{garfinkl-at-oakland.edu}
{mailto:garfinkl@oakland.edu}
\end{center}

In 1914 Einstein wrote a long review paper ``The Formal Foundation of the General Theory of Relativity'' 
on the version of general relativity theory that he had developed with Grossmann.  

\section*{\centerline
{GGR program at the APS meeting in Savannah, GA}}
\addtocontents{toc}{\protect\medskip}
\addcontentsline{toc}{subsubsection}{
\it GGR program at the APS meeting in Savannah, GA, by David Garfinkle}
\parskip=3pt
\begin{center}
David Garfinkle, Oakland University
\htmladdnormallink{garfinkl-at-oakland.edu}
{mailto:garfinkl@oakland.edu}
\end{center}

We have an exicting GGR related program at the upcoming APS April meeting in Savannah, GA.  Our Chair-Elect, Beverly Berger, did 
an excellent job of putting together this program.  At the APS meeting there will be several invited sessions of talks 
sponsored by the Topical Group in Gravitation (GGR).  

The invited sessions sponsored by GGR are as follows:\\

Issues in Quantum Gravity\\
Martin Bojowald, An Effective Framework for Quantum Cosmology\\
Bianca Dittrich, Finding the Continuum Limit of Spin Foam and Spin Net Models of Quantum Gravity\\
Parampreet Singh, Singularity Resolution in Quantum Gravity\\

Compact Binaries and Gravitational Waves: Simulations, Templates and Interpretation\\
Geoffrey Lovelace, Numerical Simulations of Merging Black Holes for Gravitational-Wave Astronomy\\
Manuel Tiglio, Reduced Order Modeling in General Relativity\\
Nicolas Yunes, Three-Hair Relations, Orbital Motion and Gravitational Waves from Neutron Star Binaries\\
 
Careers Beyond Gravitational Physics\\
Kenneth Smith, Numerical Relativity as Preparation for Industrial Data Science, a Personal Perspective\\
William Nelson, From Cosmology to Consulting\\
Michael Burka, Developing Technology Products - a Physicist's Perspective\\

Jets and Astrophysical Tests of General Relativity\\
Sera Markoff, Observationally Constraining teh Jet Power Extracted from Spinning Black Holes\\
Alexander Tchekhovskoy, GRMHD Simulations of Black Hole Accretion and Jets\\
Scott Ransom, Tests of GR using Neutron Star - White Dwarf Binaries\\

Progress Toward the Advanced Detector Era\\
Jeffrey Kissel, The Status of Advanced LIGO: Light at the End of the Tunnels!\\
Jessica McIver, Preparing to Analyze Advanced LIGO Data: from Detectors to First Observations\\
Ruslan, Vaulin, Multi-messenger Observations of Gravitational-Wave Sources in the Advanced Detector Era\\

Singularities in General Relativity\\
James Isenberg, On the Nature of Singularities in Cosmological Solutions of Einstein's Equations\\
Abhay Ashtekar, Dynamics Near Spacelike Singularities and Quantum Bounces\\
David Garfinkle, Numerical Investigations of Singularities in General Relativity\\

The Transient Gravitational Wave Sky\\
Pablo Laguna, Overview of the Transient Gravitational Wave Sky\\
Rosalba Perna, Gamma-Ray Bursts in the Gravitational Wave Era\\
Laura Cadonati, Gravitational Wave Observations Expected from the Transient Gravitational Wave Sky\\

Gravitational Waves and Nuclear Astrophysics\\
Matthew Duez, Merging ``Real'' Neutron Stars for Gravitational Waves and Electromagnetic Counterparts\\
Andrew Steiner, Neutron Star Structure, Neutron-rich Matter, and Gravitational Waves\\
Daniel Kasen, Electromagnetic Signatures of Neutron Star Mergers\\

The GGR contributed sessions are as follows:\\

Gravitational Wave Astrophysics I\\

Gravitational Wave Astrophysics II\\

Mathematical Aspects of General Relativity I\\

Mathematical Aspects of General Relativity II\\

Gravitational Wave Astrophysics III\\

Numerical Relativity in Vacuum: Methods and Simulations I\\

Quantum Aspects of Gravitation I\\

Numerical Relativity with Matter: Methods and Simulations I\\

Quantum Aspects of Gravitation II\\

Numerical Relativity in Vacuum: Methods and Simulations II\\

Numerical Relativity with Matter: Methods and Simulations II\\

Tests of General Relativity and Gravitation\\

Gravitational Waveform Modeling\\

Alternate Theories of Gravity\\

Approximations to General Relativity\\

Gravitational Wave Experiment\\

Frontiers in Gravitation\\

\vfill\eject

\vfill\eject
\section*{\centerline
{GR20/Amaldi10 Conference held at Warsaw, Poland}}
\addtocontents{toc}{\protect\medskip}
\addtocontents{toc}{\bf Conference reports:}
\addcontentsline{toc}{subsubsection}{
\it GR20/Amaldi10 Conference held at Warsaw, Poland, 
by Abhay Ashtekar}
\parskip=3pt
\begin{center}
Abhay Ashtekar, IGC, Penn State 
\htmladdnormallink{ashtekar-at-gravity.ps.edu}
{mailto:ashtekar@gravity.ps.edu}
\end{center}

The GR conferences are tri-annual while the Amaldi conferences are
bi-annual. So they can overlap every 6 years. They did so in 2013
and a joint conference was held at Warsaw, Poland, from July 7th to
13th. It drew some 850 participants, probably the largest number
either series has attracted so far. As a result, it was an
impressive and stimulating gathering, covering a wide range of
topics including geometric analysis and numerical relativity,
relativistic astrophysics and cosmology, gravitational wave science
and other gravitational experiments, and quantum aspects of gravity.

Since this joint conference was held on the eve of the first century
of general relativity, it was a major event with historic
significance for our field. Furthermore, it took place 51 years
after GR3, which was also held at (Jablona near) Warsaw. Therefore,
in the opening ceremony the organizers held a special session with
two talks, one on the participants and atmosphere at GR3, a landmark
conference in our field, and another on scientific advances over the
past half century, comparing what we knew at GR3 with what we now
know. The opening ceremony also included a prize session. Three
thesis prizes were presented: \\
i) The GWIC prize, of the Gravitational Wave International
Committee, was awarded to \emph{Dr. Paul Fulda} who carried out his
thesis work at the University of Birmingham; \\
ii) The J\"urgen Ehlers prize for mathematical and numerical general
relativity, sponsored by Springer, publisher of the Society Journal,
was awarded to \emph{Dr. Aseem Paranjape} who did his thesis work at
the Tata Institute for Fundamental Research in Mumbai;\\
iii) The Bergmann-Wheeler prize in the broad area of quantum
gravity, sponsored by the Institute of Physics, publisher of CQG,
was awarded to \emph{Dr. Aron Wall} who carried out his thesis work
at the University of Maryland.\\
In addition, the first IUPAP Young Scientist Prize for General
Relativity and Gravitation was presented to \emph{Dr. Lisa
Barsotti}. It replaces the Xanthopoulos prize, awarded by the GRG
Society since 1991, and carries with it an IUPAP Gold Medal in
addition to a monetary award.

The scientific part of the conference was held in the historic part
of Warsaw University near the city center. The local organizing
committee Chaired by Prof. Jerzy Lewandowski did an outstanding job,
looking after every detail. Furthermore, they managed to make
special arrangements with their university administration to keep
the registration fee exceptionally low, setting an excellent example
for future conferences in these series. The international scientific
organizing committees were Chaired by Prof. Bala Iyer for GR20, and
Prof. Sheila Rowan for Amaldi10. They succeeded in creating a truly
stimulating conference with 20 plenary lectures and 25 parallel
sessions. In addition, there was an evening public lecture in which
Prof. Carlo Rovelli provided a lucid explanation of the main ideas
behind some of the forefront research on quantum gravity using an
easy to grasp, intuitive format.

The scientific program and slides from plenary and parallel sessions
are available on the web page
http://gr20-amaldi10.edu.pl/index.php?id=1  Furthermore, since the
conference represented a historic landmark, articles based on
plenary lectures and summaries of workshops will appear as a
\emph{Topical Collection GR20/Amaldi10} of the Society journal
General Relativity and Gravitation, with guest editors Bala Iyer,
Jerzy Lewandowski and Sheila Rowan. These reports not only summarize
the conference but also present a good overview of where the field
is at the end of the first century of general relativity. Therefore,
here I will attempt to provide just an overall impression through a
few glimpses of the scientific program.

On the mathematical and numerical side, Prof. Piotr Bizon presented
his striking result that the anti-de Sitter space-time, is
\emph{unstable} in full, non-linear general relativity although it
was known to be perturbatively stable. This is surprising both
because it has been known for some time that de Sitter space-time
and Minkowski space-time are non-linearly stable and because the
anti-de Sitter space-time is assumed to represent the `ground state'
in the ADS/CFT correspondence. On the astrophysical side, Prof.
Priya Natarajan summarized recent advances in the understanding of
the growth of supermassive black holes. Discoveries of quasars at
high red-shifts ($z>6$) imply that there were black holes of a
billion solar masses already when the universe was only a couple of
billion years old. The talk discussed mechanisms for black hole seed
formation at high $z$, summarized the challenges faced by models in
matching current observational data, and concluded with a discussion
of observational prospects for discriminating between models. Prof.
Georges Efstathiou summarized the cosmological results from the
first 15 months of data of the Planck mission. The results agree
spectacularly well with the basic six parameter, $\Lambda$CDM
cosmology, with a definitive departure from complete scale
invariance, and no evidence of significant non-Gaussianities. But
there were also some anomalies, such as a hemispherical asymmetry in
power, which, if statistically significant, will guide future
theoretical work. There were several talks on the gravitational wave
science since we are now at the threshold of an exciting epoch in
this area. Prof. Marie Anne Bizouard summarized the results to date
from the LIGO-VIRGO-GEO collaboration. In particular, the
\emph{non-detection} of gravitational waves at the current
sensitivity has ruled out the conjectured binary coalescence
associated with the gamma ray burst sources GRB 070201 and GRB
051103, at a 99\% confidence level in the galaxy M31 and at a 98\%
confidence level in M81. She also discussed the expected enhancement
in the science reach in the next 2-3 years, once the the advanced
detectors start taking data. Prof. Stefano Vitale reviewed the
concepts underlying space-based detectors and summarized the status
of LISA pathfinder which will be launched by the European Space
Agency in 2015. Prof. Maura McLaughlin summarized how pulsar timing
arrays can be used as a powerful tool to detect stochastic
gravitational waves. Since a likely source of these waves are an
ensemble of supermassive black hole binaries, pulsar timing arrays
can be used to constrain models of galaxy formation. The expected
discovery of many new millisecond pulsars should significantly
improve the sensitivity of these measurements. This avenue is also
expected to lead to a detection of gravitational waves during the
decade starting 2015. Finally, on quantum aspects of gravity, Prof.
Rajesh Gopakumar chaired a stimulating, 3 hour, joint session on the
quantum mechanics of black hole evaporation. It featured 7 talks
with lively exchange among the speakers, sharp questions from the
audience and informative responses from the panel. Although there is
agreement among experts on many of the core issues, even after 4
decades of debate, key differences still remain. This is because one
does not yet have a satisfactory quantum gravity theory to
decisively tell us what really happens to the classical singularity.
The session was successful in that it brought out core differences
between various approaches, thereby bringing the key conceptual
issues into sharper focus.

The conference dinner was held on Wednesday in a historic, sumptuous
venue and featured after-dinner remarks by Prof. Ezra T. Newman. As
per tradition, the business meeting of the GRG Society was held in
the evening on Thursday. Prof. Gary Horowitz was elected President
and Dr. Beverly Berger was re-elected Secretary and Treasurer. As
per the bylaws of the Society, Prof. Malcolm McCallum, the past
President, will now serve as Deputy President. A complete list of
the newly elected members of the international committee can be
found at http://www.isgrg.org/committee.php  The society also
inducted 15 new Fellows. Five of them are former Presidents of the
Society, and, as per policy approved by the International Committee
in 2008, five are below the age of 45. A complete list of the
Fellows can be found at http://www.isgrg.org/fellows.php   After the
conference, in consultation with the workshop Chairs and the Chair
of the Scientific Committee, the President of the GRG Society
awarded four S. Chandrasekhar prizes (sponsored by World Scientific
publishers) for the best post-doc presentations in the parallel sessions to:\\
\indent {\it Charles Melby-Thompson} (session A3), {\it Alexandre Le
Tiec} (B4 and D4), {\it John Veitch} (C2), and, {\it Eric Perlmutter} (D2); \\
and 10 Hartle awards (endowed by Prof. James B. Hartle) for the best
student presentations in the parallel sessions:\\
\indent {\it Majid Abdelqadet} (A1), {\it Christopher Berry} (B1),
{\it Stephanie Erickson} (B2),  {\it Sydney Chamberlain} (C1), {\it
Oliver Gerberding} (C5 and C8), {\it Giuliana'Russano} (C9), {\it
Antonia Zipfel} (D1), {\it Gavin'Harnett}
(D2), {\it Lisa Glaser} (D3), and, {\it Valentina'Baccetti} (D4).\\

Finally, two satellite schools were held in conjunction with this
conference, both highly successful. The first was on Quantum and
Mathematical Gravity held at Zakopane, a resort near Krakow, (see:
http://th-www.if.uj.edu.pl/school/2013/ ), and the second was on
Gravitational Waves, held at the Banach Center in Warsaw (see,
http://bcc.impan.pl/13Gravitational/ ).\\

All in all, GR20/Amaldi 10 and the associated academic initiatives
were extremely well organized and stimulating international events
for our field. The next GR conference, GR21 will be held in the
summer of 2016 at Columbia University, New York and Amaldi11 will be
held at Gwangju, Korea in the summer of 2015.

\vfill\eject
\section*{\centerline
{School and workshop on quantum gravity in Brazil}}
\addtocontents{toc}{\protect\medskip}
\addcontentsline{toc}{subsubsection}{
\it School and workshop on quantum gravity in Brazil, 
by Jorge Pullin}
\parskip=3pt
\begin{center}
Jorge Pullin, Louisiana State University 
\htmladdnormallink{pullin-at-lsu.edu}
{mailto:pullin@lsu.edu}
\end{center}

School and workshop on quantum gravity in Brazil

by Jorge Pullin, Louisiana State University

There was a School on Quantum Gravity held at the ICTP-SAIFR (International
Centre for Theoretical Physics, South American Institute for Fundamental Research)
at Sao Paulo, September 2nd-9th, followed by the by now traditional
workshop "Quantum Gravity in the Southern Cone" at the nearby beach resort
of Maresias September 11th-14th 2013.

At the school, lectures were delivered by Abhay Ashtekar, Juan Maldacena,
Alejandro Perez, Martin Reuter, John Schwarz, Erik Verlinde and Raul Abramo,
covering topics ranging from String Theory, Loop Quantum Gravity, Asymptotic
Safety, AdS/CFT and cosmology and entropic gravity.

The Quantum Gravity in the Southern Cone Workshop has been ocurring
regularly every three years and this was its sixth edition. There were talks
by Gerardo Aldazabal, Jorge Alfaro, Abhay Ashtekar, Max Bañados, Nathan
Berkovits, Steven Carlip, Marc Casals, Diego Correa, Bianca Dittrich,
Jose Edelstein, Gaston Giribet, Aleksandr Pinzul, Martin Reuter, Jorge
Russo, Kelly Stelle, Daniel Sudarsky, Diego Trancavelli and Olivera Miskovic.
Again, the talks covered a wide range of topics in various approaches
to quantum gravity.

The next edition of Quantum Gravity in the Southern Cone will take place
in 2016 in Uruguay.

\end{document}